\documentclass[preprint,nofootinbib,preprintnumbers,amsmath,amssymb]{revtex4}

\usepackage{graphicx}
\usepackage{dcolumn}
\usepackage{bm}

\RequirePackage{xspace}
\usepackage{relsize}

\def\Kbar  {\kern 0.2em\overline{\kern -0.2em K}{}\xspace}

\def\Bbar    {\kern 0.18em\overline{\kern -0.18em B}{}\xspace}

\def\Qbar    {\kern 0.08em\overline{\kern -0.08em Q}{}\xspace}

\newcommand{\mev}{\ensuremath{\mathrm{\,Me\kern -0.1em V}}\xspace}
\newcommand{\mevc}{\ensuremath{{\mathrm{\,Me\kern -0.1em V\!/}c}}\xspace}
\newcommand{\mevcc}{\ensuremath{{\mathrm{\,Me\kern -0.1em V\!/}c^2}}\xspace}
\newcommand{\gev}{\ensuremath{\mathrm{\,Ge\kern -0.1em V}}\xspace}
\newcommand{\gevc}{\ensuremath{{\mathrm{\,Ge\kern -0.1em V\!/}c}}\xspace}
\newcommand{\gevcnospace}{\ensuremath{{\mathrm{\,Ge\kern -0.1em V\!/}c}}}
\newcommand{\gevcc}{\ensuremath{{\mathrm{\,Ge\kern -0.1em V\!/}c^2}}\xspace}
\newcommand{\be}{\begin{equation}}
\newcommand{\ee}{\end{equation}}
\newcommand{\benn}{\begin{equation*}}
\newcommand{\eenn}{\end{equation*}}
\newcommand{\bea}{\begin{eqnarray}}
\newcommand{\eea}{\end{eqnarray}}

\def\beq{\begin{equation}}
\def\eeq{\end{equation}}
\def\bea{\begin{eqnarray}}
\def\eea{\end{eqnarray}}

\newcommand{\lsim}{ \mathop{}_{\textstyle \sim}^{\textstyle <} }

\newcommand{\kev}{{\rm keV}}

\usepackage{color}

\begin{document}

\preprint{UCI-HEP-TR-2011-15}

\title{LHC Bounds on Interactions of Dark Matter\\[0.25cm]}

\author{Arvind~Rajaraman, William~Shepherd, Tim M.P.~Tait and Alexander M.~Wijangco
\\[0.5cm]
{\small {\it Department of Physics and Astronomy,}}
\\
{\small {\it University of California, Irvine, California 92697}}}
\date{\today~\\[2cm]}

\begin{abstract}
We derive limits on the interactions of dark matter with
quarks from ATLAS null searches for jets + missing energy
based on $\sim 1~{\rm fb}^{-1}$ of integrated luminosity, 
using a model-insensitive effective theory framework.
We find that the new limits from the LHC significantly extend limits previously
derived from CDF data at the Tevatron.
Translated into the parameter space of direct searches, these limits are
particularly effective for $\sim$~GeV mass WIMPs.  
Our limits indicate tension with isospin
violating models satisfying minimal flavor violation which attempt to reconcile the purported CoGeNT excess with
Xenon-100, indicating that either a light mediator or nontrivial flavor structure for the 
dark sector is necessary for a viable reconciliation of CoGeNT with Xenon.
\end{abstract}

\maketitle
\section{Introduction}
\label{sec:intro}

The evidence for the existence of dark matter in the universe is 
overwhelming\,\cite{Komatsu:2010fb},
and models to incorporate dark matter into our understanding of the fundamental
physics of the universe are myriad. Astrophysical observations tell us nothing
about the mass of the dark matter particle or whether it interacts 
with the Standard Model (SM) particles in any way
other than gravitationally. Models range in masses from \kev~to the GUT scale,
and in coupling strength from slightly weaker than QCD couplings to purely
gravitational interactions. The most popular models are driven by the
WIMP(less) miracle\,\cite{Feng:2008ya}, suggesting that the dark matter particle 
relic density should naturally be set by the thermal history of the universe and
favoring a
ratio of the mass and coupling strength.  Such
dark matter candidates naturally
appear in extensions of the Standard Model which are designed to address other
theoretical issues, most notably the gauge hierarchy problem. Since WIMPs have
fairly large couplings to SM fields to explain their relic density, it is
possible to search for them interacting directly with normal matter,
annihilating into normal matter, or being produced at high energy colliders.

Any WIMP which produces a signal in one of these searches would 
naively be expected
to be seen in others as well, as a single coupling could be visible to 
all of them. Each type of experiments has a
particular set of strengths and weaknesses associated with its ability to
discover or exclude various models of dark matter. Direct detection experiments
have a signal that is strongly peaked at very low energies,
making it hard to distinguish from background effects and causing detector
thresholds to be particularly troublesome when light candidate particles are
considered. Indirect detection searches for dark matter annihilation products
are able to observe locations which have much higher local densities of dark
matter than our solar system, but then must contend with large astrophysical
background uncertainties. Colliders have a fixed amount of energy available
to them in the collisions
(and do not take advantage of the dark matter already present in the galactic
halo), and are thus unable to produce dark matter of
very large mass, but have exceptional sensitivity to low mass WIMPs, which are
ill constrained by the other two techniques. Any signal seen at
colliders may be due to other new physics than dark matter, so astrophysical
confirmation will be critical to being able to make robust claims regarding
dark matter at colliders. However, colliders are able to make strong exclusion 
statements in the event of no signal\,
\cite{Birkedal:2004xn,Beltran:2008xg,Cao:2009uw,Beltran:2010ww,Shepherd:2009sa,Goodman:2010yf,Goodman:2010ku,Bai:2010hh,Fox:2011fx,Fortin:2011hv,Kamenik:2011nb,Burgess:2000yq,Kanemura:2010sh,Cheung:2010zf,Wang:2011sx}.

Currently there is much interest in light WIMP models, with masses of
order $\lsim10\gev$, motivated in large part by experimental results from the
CoGeNT collaboration\,\cite{Aalseth:2010vx,Aalseth:2011wp} 
which can be explained by such a WIMP and appear to be
tantalizingly close to the parameter space favored by
a dark matter interpretation of the longstanding DAMA annual
modulation signal\,\cite{Bernabei:2010mq,Petriello:2008jj,Feng:2008dz}. 
As the CoGeNT collaboration has recently reported
that they also see annual modulation in their data\,\cite{Aalseth:2011wp}, 
these results have only grown more interesting.
These putative signals are, however, in significant
tension with negative results from the Xenon 100 \cite{Aprile:2010um} 
and CDMS-II \cite{CDMS-Si,Ahmed:2009zw,Ahmed:2010wy} collaborations,
and the modulation exhibits an unexpected dependence on the recoil
energy of the scattered nucleus 
\cite{Schwetz:2011xm,Fox:2011px,Farina:2011pw}.

In this work we extend previous 
studies\,\cite{Goodman:2010yf,Goodman:2010ku,Bai:2010hh,Fox:2011fx,Fan:2010gt} 
which use the framework of effective field theory to construct models of 
dark matter and constrain them from collider searches. These models make
specific predictions for other dark matter searches as well, and allow the
collider constraints to be drawn on a direct detection plane. Similarly,
constraints from indirect searches can be interpreted in these models on the
plane of direct detection\,\cite{Goodman:2010qn,Cheung:2010ua,Buckley:2011kk}.
We enlarge our previous set of
effective theories to allow couplings to only one type of quarks at a time.
This allows for the inclusion of effects which distinguish between quark charge
in the model-independent framework which we previously presented
and are more representative of the range of possible couplings present in
models with minimal flavor violation (MFV) \cite{Buras:2000dm}. 
In particular, the dependence on
$\tan \beta$ expected in type-II 
two Higgs doublet extensions of the SM
(such as in the Minimal Supersymmetric Standard Model) 
can be easily represented in this set of models, in contrast to
our previous work.

A recent proposal\,\cite{Feng:2011vu,Chang:2010yk}
suggests that dark matter interactions may be sensitive to the specific
proton and neutron content of the nucleus with which it is scattering, 
rather than just the net baryon number (the mass of the nucleus). 
For a WIMP whose couplings satisfy $\lambda_n / \lambda_p \sim -0.7$,
one obtains 
consistency between the negative results of the Xenon collaboration and the
putative signal seen at CoGeNT by largely canceling the coupling to xenon
nuclei.  This parameter point has the additional feature that it shifts the
DAMA target region such that it moves from being close to but inconsistent
with the CoGeNT signal, to a situation where CoGeNT and DAMA are
fit by consistent choices of parameters.
In a short time, many models predicting or utilizing this
``isospin-violating"
mechanism have appeared in the 
literature\,\cite{Kang:2011wb,DelNobile:2011je,Gao:2011ka,Frandsen:2011ts,Gao:2011bq}.

This article is organized as follows: In section\,\ref{sec:eft} we discuss the
effective field theory modeling of WIMP-SM interactions, in 
section\,\ref{sec:collider} we calculate bounds on the strength of dark matter
interactions using collider data and present future reach for the LHC, in
section\,\ref{sec:direct} we discuss the impact of these bounds on direct
detection signals, and in section\,\ref{sec:conclusion} we present our conclusions.

\section{Model Description}
\label{sec:eft}

In formulating our constraints on dark matter from collider searches we assume
that the dark matter candidate is the only new particle which is accessible
at the relevant experiments and that dark matter is a SM gauge singlet.
Under these assumptions, only non-renormalizable couplings are possible 
between dark matter and the SM fields. 
We therefore focus on the operators which are of the
lowest dimensionality, as these will give the strongest signals at energies
below the scale which characterizes the interactions.

As with any effective field theory, the models of dark matter we construct in
this way are only applicable below some cutoff scale where other new physics
becomes relevant and renormalizability is regained. This cutoff is approximately
at the mass of the lowest-lying state which is integrated out in the effective
theory. This is related to the scale suppressing the higher-dimensional
operators and the couplings of the fundamental theory as
\beq
M_*\sim\frac{M_\Phi}{g_\Phi},
\eeq
where $\Phi$ is the field which has been integrated out to give the interaction
whose strength is parametrized by $M_*$. Note that this relation tells us that
below a certain value of $M_*$ it is not possible to have a perturbative
completion of the theory involving exchange of
particles whose masses are all larger than the WIMP
mass; we discard results in such regions as it is clear there is no
perturbative UV completion of the effective theory in this regime\,\cite{Goodman:2010yf}.

In this work our primary focus is on the effect that isospin violation can have
on collider constraints on dark matter, so we will specialize to the case of a
Majorana WIMP, as constraints on the isospin violating 
couplings from colliders are not expected to depend sensitively on the 
nature of the dark matter candidate 
\cite{Goodman:2010yf,Goodman:2010ku,Bai:2010hh}. 
As we are particularly interested in relating to
direct detection, we focus on
couplings of dark matter to quarks. Gluon couplings are also interesting for
direct detection, but they are not able to differentiate between states of
different isospin. We therefore do not consider couplings of dark matter to
gluons in this work.

We construct all of the lowest-dimension operators that couple 
dark matter and quarks consistent with
MFV, which helps ensure that the models which we produce
are not in conflict with flavor physics observables \cite{Kamenik:2011nb}. 
This amounts to the 
assumption that any term which breaks $SU(2)_L$ of the SM must do so through
the SM Yukawa couplings, leading to the suppression 
by the quark mass of any operator
which flips the quark chirality. The leading operators are of the form
\beq
\label{eq:operators}
L_{Eff} = G_\chi~\bar\chi\Gamma_\chi\chi~\bar q\Gamma_qq 
\eeq
where
\beq
\Gamma_{\chi,q}\in\{1,\gamma^5,\gamma^\mu,\gamma^\mu\gamma^5,\sigma^{\mu\nu}\}.
\eeq

\begin{table}[tdp]
\begin{center}
\begin{tabular}{|c|c|c|c|c|}
\hline
   Name  & $G_\chi$ & $\Gamma_\chi$ & $\Gamma_q$ \\
\hline
M1 & $m_q/2M_*^3$  &  $1$                     & $1$ \\
M2 & $im_q/2M_*^3$  &  $\gamma_5$                     & $1$ \\
M3 & $im_q/2M_*^3$  &  $1$                     & $\gamma_5$ \\
M4 & $m_q/2M_*^3$  &  $\gamma_5$                     & $\gamma_5$ \\
M5 & $1/2M_*^2$  &  $\gamma_5\gamma_\mu$                     & $\gamma^{\mu}$ \\
M6 & $1/2M_*^2$  &  $\gamma_5\gamma_\mu$                     & $\gamma_5\gamma^{\mu}$ \\
\hline
\end{tabular}
\caption{The list of the effective operators defined in Eq.\,(\ref{eq:operators}).
\label{tab:ops}}
\end{center}
\end{table}

Any other combination of bilinears are equivalent to a linear combination of this set 
through Fierz identities. Note that any Lorentz indices in $\Gamma_\chi$ 
must be contracted with indices
in $\Gamma_q$ to preserve Lorentz invariance. Thus our models contain no
tensor terms, because it vanishes for Majorana particles
and the alternatives are higher order in derivatives, and thus
more suppressed in low energy reactions. The MFV assumption
requires us to scale quark bilinears with no Lorentz indices by the quark mass,
and to have no relative scaling between the couplings for different quarks in
bilinears carrying a Lorentz index. However, we still have
two independent coefficients for each operator structure associated with up-
and down-type quark couplings, which are not constrained relative to each other
by MFV.

The list of all operator Lorentz structures we consider are
presented in Table\,\ref{tab:ops}. Note that the cases of up- and down-type
couplings are distinguished in our notation by a trailing u or d on the
designator of the Lorentz structure.  For example, operator M1u 
corresponds to
\bea
{\cal L}_{\rm M1u} & = & \frac{1}{2 M_*^3}~  \overline{\chi} \chi 
\sum_{q=u,c,t} m_q~ \overline{q} q .
\eea

\section{Collider Searches}
\label{sec:collider}

We constrain the
operators by simulating the production of a pair of WIMPs and jets at colliders,
\beq
pp (p\bar{p}) \to \chi \chi + {\rm jets}
\eeq
As the WIMPs are invisible to the particle detectors, such a process would appear 
as a combinations of jets and missing energy. We estimate efficiencies for
the signal to pass analysis cuts (outlined below) based on simulations using 
Madgraph 4.5.0, with showering and detector simulation performed by the 
Madgraph Pythia-PGS 2.8 package \cite{Alwall:2007st,Sjostrand:2006za,PGS}. 
The dominant standard model background for such a signal is $Z$ + jets, 
where the $Z$ boson then decays to a pair of neutrinos. 
The next largest background is $W$ + jets, where the $W$ 
decays into a neutrino and a charged lepton which is mistagged to be a jet or
lost \cite{Aaltonen:2008hh,CDF,AtlasConf2011}. 

We assume only one Lorentz structure is dominant at a time, 
and constrain each by assuming the others do not contribute to the cross section. 
Since the coupling of models with scalar Lorentz structures are proportional to 
quark mass, the cross sections from down-type operators are enhanced by the 
bottom quark mass (though moderated by the $b$ parton distribution function),
resulting in stronger bounds on operators M1d--M4d compared to
M1u--M4u.
For models with vector Lorentz structure, the parton distribution functions 
are the dominant difference between the up-type and down-type
operators, resulting in comparatively stronger constraints upon the up-type 
couplings. 

\begin{figure}[t]
\includegraphics[width=17.0cm]{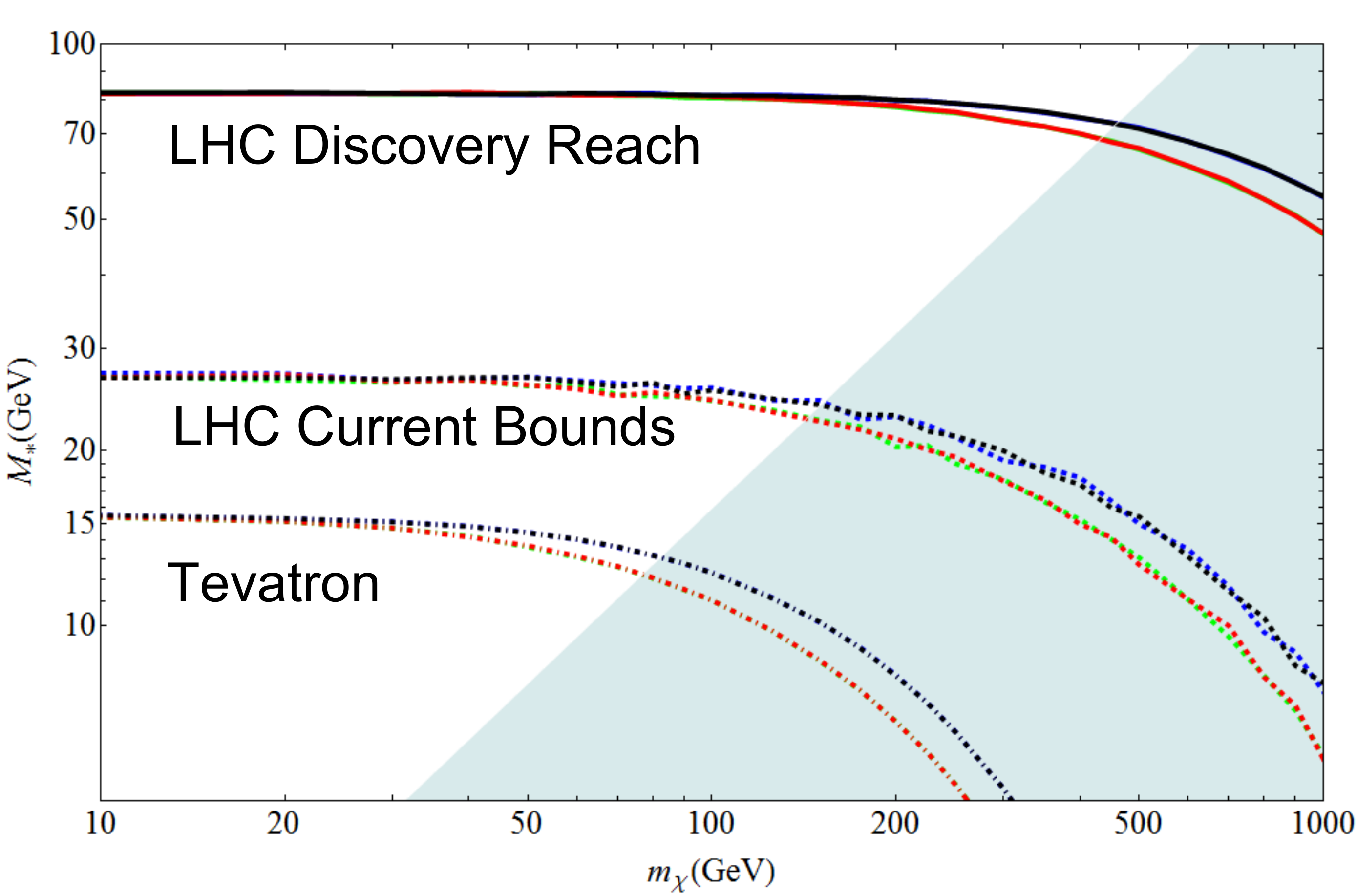}
\caption{\label{fig:csd} 
The collider bounds on the down-type quark operators with scalar Lorentz 
structures. Operators M1d, M2d, M3d, M4d, are in red, blue, green, and black 
respectively. The dashed-dotted, dashed, and solid lines are the Tevatron 
constraints, LHC constraints, and LHC discovery reach. The shaded region is where 
the effective theory breaks down. Models M1d and M3d are largely degenerate, as 
are models M2d and M4d.
}
\end{figure}

\begin{figure}[t]
\includegraphics[width=17.0cm]{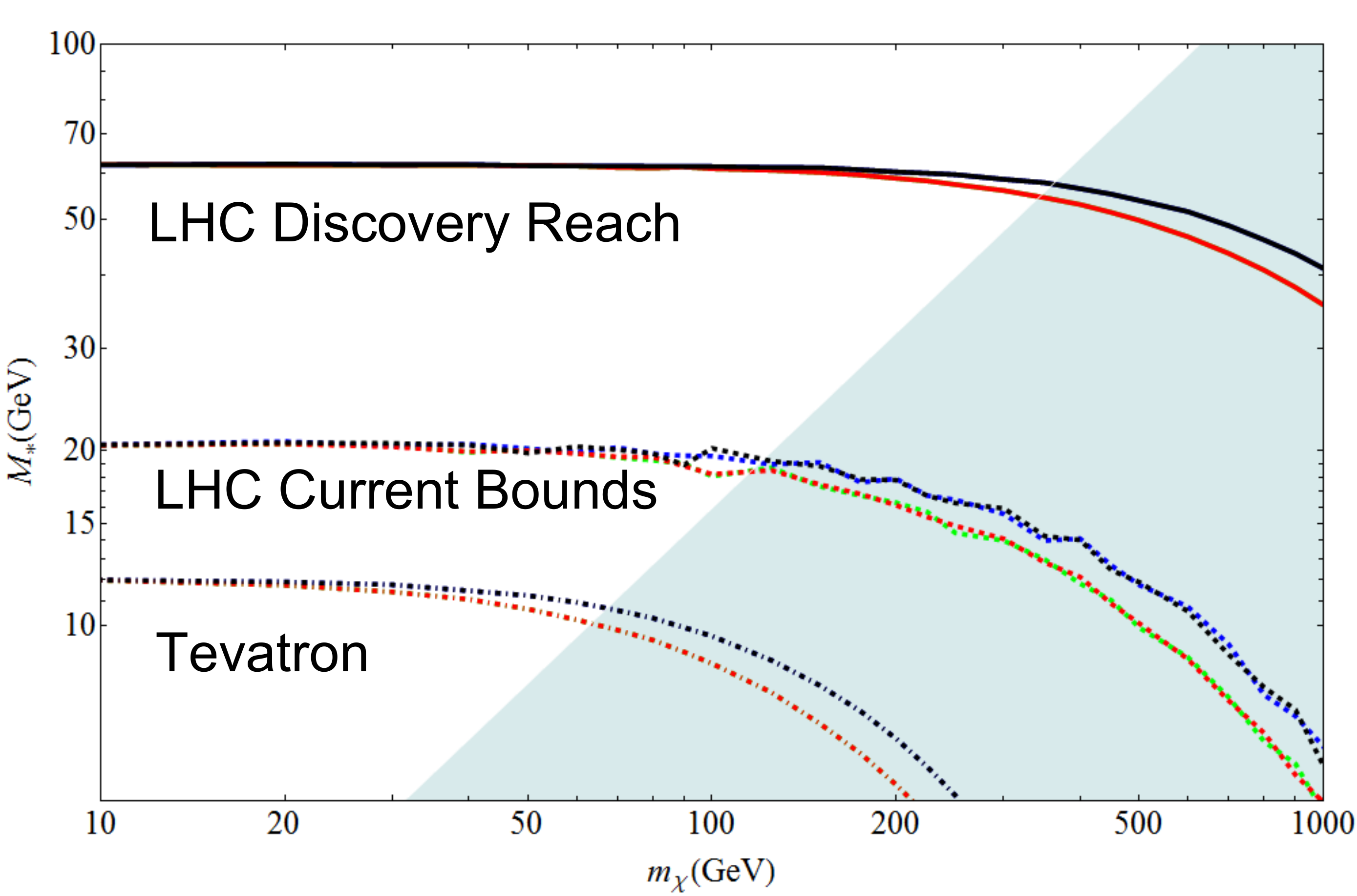}
\caption{\label{fig:csu} 
The same as Fig. \ref{fig:csd}, for up-type quark 
operators M1u, M2u, M3u, and M4u.
}
\end{figure}

\begin{figure}[t]
\includegraphics[width=17.0cm]{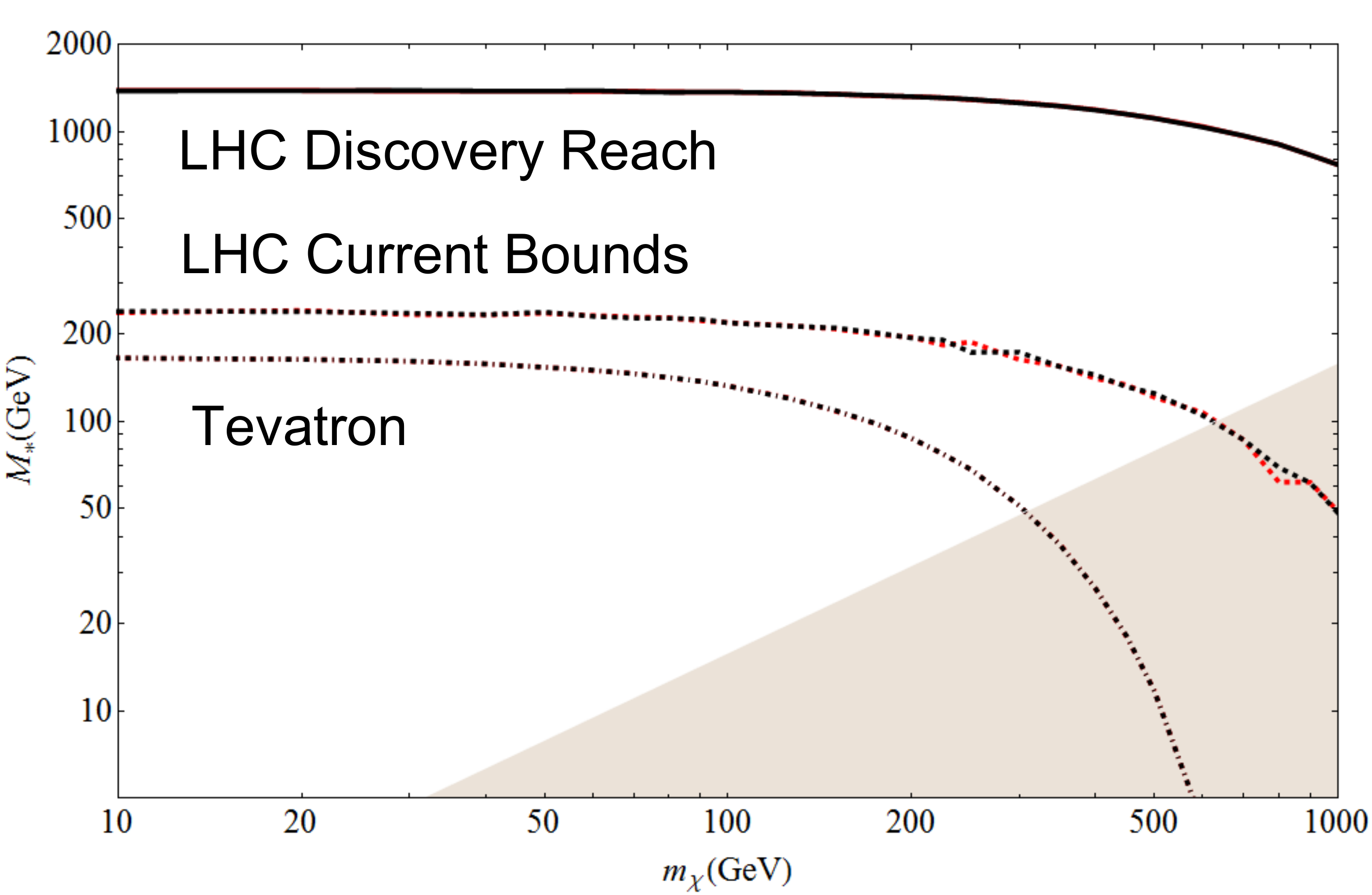}
\caption{\label{fig:cvd} 
The collider bounds on the down-type quark coupling operators mediated by a heavy scalar. Models M5d, M6d are in red, black respectively. The dashed-dotted, dashed, and solid lines are the Tevatron constraints, LHC constraints, and LHC discovery reach. The shaded region is where the effective theory breaks down. Models M5d and M6d are largely degenerate.
}
\end{figure}

\begin{figure}[t]
\includegraphics[width=17.0cm]{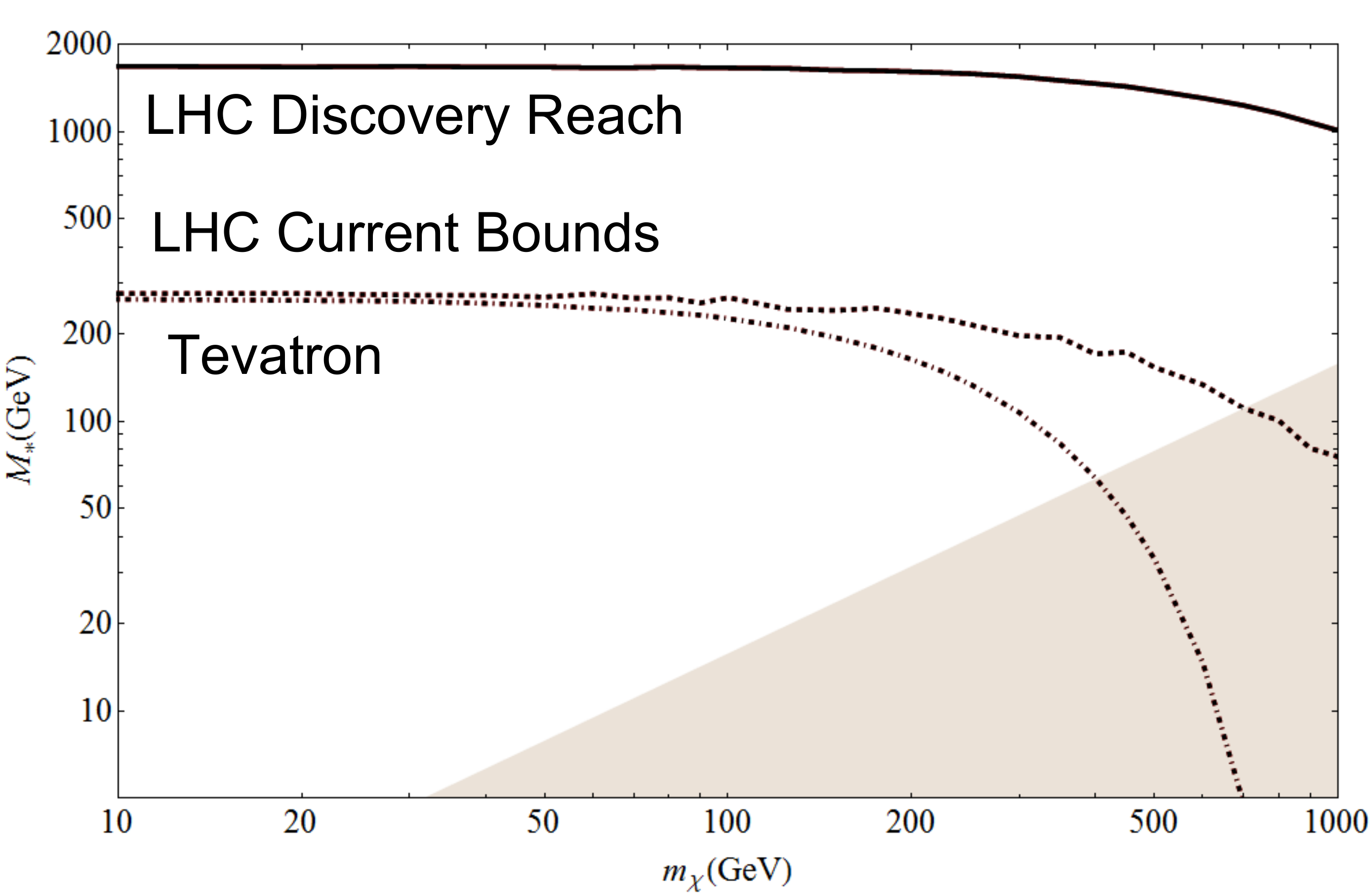}
\caption{\label{fig:cvu} 
The same as Fig. \ref{fig:cvd}, although now the up-type quark coupling operators M5u and M6u are displayed.}
\end{figure}

\subsection{Tevatron Constraints}

The CDF collaboration has reported null results for a mono-jet search 
based on about 1 fb$^{-1}$ of Tevatron run II data \cite{CDF}, 
constraining the size of additional contributions to missing energy + jets.  
The analysis selects events which have
missing transverse momentum $\not \hspace*{-0.12cm} E_T\ >$ 80 GeV together 
with a leading jet whose transverse momentum is $p_T\ >$ 80 GeV.  A 
second jet with $p_T\ <$ 30 GeV is allowed, and any 
subsequent jets must have $p_T\ <$ 20 GeV. 
In a sample size of 1 fb$^{-1}$, CDF found 8449 events while the Standard Model 
prediction was 8663 $\pm$ 332 events. To be within 2$\sigma$ of these results, 
the accepted cross section of new physics can be at most 0.664 pb. 
In Figures \ref{fig:csd} - \ref{fig:cvu}, we translate the
cross section limit into one on $M_*$ for each operator, as a function of the
dark matter mass.

\subsection{LHC Constraints}

The ATLAS Collaboration has very recently released the results of a search
for anomalous production of jets and missing energy 
at $\sqrt{s}$ = 7 TeV with an integrated luminosity of $1.00$~fb$^{-1}$
\cite{AtlasConf2011}. 
Events with $\not \hspace*{-0.12cm} E_T >$ 120 GeV and containing
a leading jet with $p_T >$ 120 GeV and $|\eta| <$ 2 were selected.
A second jet with $p_T <$ 30 GeV and $|\eta| < 4.5$ was allowed. 
15740 events were observed, to be compared with an expected 
15100$\pm$170(stat.)$\pm$680(syst.). 
This excludes an effective cross section of 1.7 pb, which we map to 
constraints upon M$_*$ in Figures \ref{fig:csd} - \ref{fig:cvu}.

\subsection{LHC Future Reach}

We also investigate the 5$\sigma$ discovery reach of such operators, using the 
analysis done in \cite{Vacavant:2001sd}, which considered the LHC running 
at $\sqrt{s}$ = 14 TeV and with an integrated luminosity of 
100~fb$^{-1}$. Events with missing 
$\not \hspace*{-0.12cm} E_T >$ 500 GeV and at least one a jet with 
$p_T >$ 500 GeV were considered, but no secondary jet rejection
cut was employed.  Events with isolated charged leptons were rejected.
Ref \cite{Vacavant:2001sd} predicts a Standard Model background of 
about $B = 3 \times 10^4$ events for this integrated luminosity.  
We determine the discovery reach by requiring that the significance of the 
new physics signal $S$ passing the cuts satisfy $S/\sqrt{B} \geq 5$ and plot the
resulting region in Figures \ref{fig:csd} - \ref{fig:cvu}.

\section{Direct Detection}
\label{sec:direct}

Our effective theory allows one to translate the collider bounds into the parameter
space of direct detection experiments.  In the non-relativistic limit, only
operators M1d, M1u, M6d, and M6u mediate unsuppressed scattering
cross sections with nucleons.  In terms of $M_*$, the resulting
cross sections are
\bea
\sigma^{p,n; SD} 
= \frac{4 \mu^2_\chi}{\pi}
\left(\frac{\Delta^{p,n}_u}{M_{*,M6u}^2}+
\frac{\Delta^{p,n}_d+\Delta^{p,n}_s}{M_{*,M6d}^2}\right)^2,\\
\sigma^{p,n; SI} 
= \frac{\mu^2_\chi}{\pi}
\left(\frac{\sum_uf^{p,n}_u}{M^3_{*,M1u}}
+\frac{\sum_df^{p,n}_d}{M^3_{*,M1d}}\right)^2,
\eea
where we have adopted the values \cite{Belanger:2008,Giedt:2009mr},
\begin{align}
\Delta_u^p=0.78,&\qquad\qquad\qquad \Delta_d^p=-0.48&\Delta_s^p=-0.15\nonumber\\
f_u^p=0.023,&\qquad\qquad\qquad f_d^p=0.033,&f_s^p=0.05,\nonumber\\
f_u^n=0.018,&\qquad\qquad\qquad f_d^n=0.042,&f_s^n=0.05,\nonumber\\
&\qquad\qquad\qquad f_{c,b,t}^{p,n}=0.066,
\end{align}
and the neutron and proton spin fractions are related by isospin symmetry.

In constructing models which have particular isospin behavior with respect to
protons and neutrons in spin-independent scattering we solve the equation
\bea
\frac{\lambda_n}{\lambda_p}
& = & \frac{\sum_df^p_d}{\sum_uf^p_u}~
\frac{M^3_{*,M1u}}{M^3_{*,M1d}},
\eea
where the ratio of neutron to proton couplings is taken as input and we
calculate the ratio of suppression scales. The models are then constrained at
colliders by noting that there is no interference at leading order between the
up- and down-type couplings, which allows us to directly sum the signal cross
section from each to find the total cross section expected for a given operator
strength.

\begin{figure}[t]
\includegraphics[width=17.0cm]{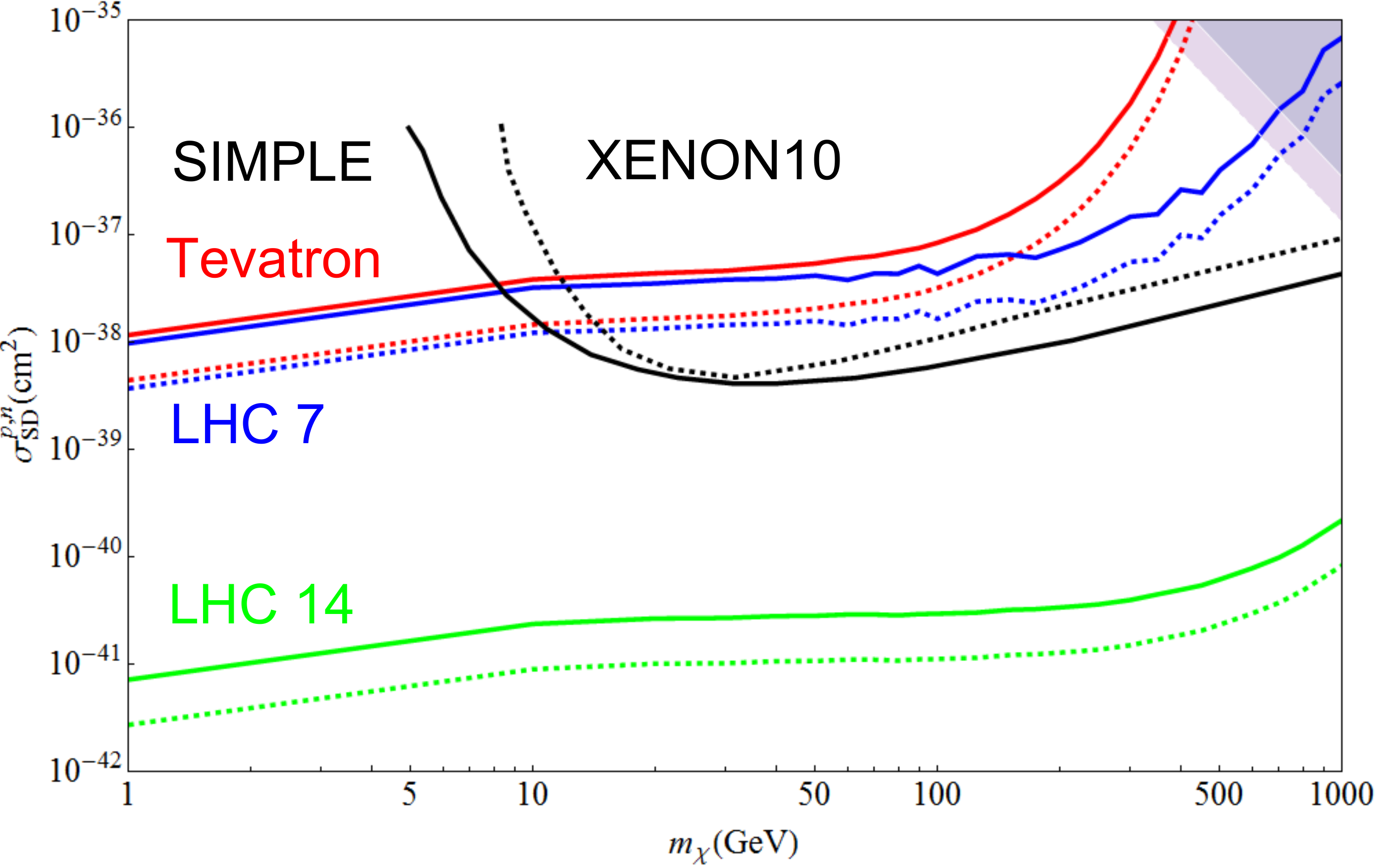}
\caption{\label{fig:dsdu} 
Spin-dependent nucleon scattering cross section assuming only the
up-type quark operator M6u is present. 
The red and blue lines are the constraints from the Tevatron search and 
7 TeV LHC search. 
The green lines are the 14 TeV LHC discovery reach. 
The solid lines are the proton coupling cross section and the dotted lines are 
the neutron coupling cross section. The dashed black line is the Xenon 10 
constraint on the neutron cross section  \cite{Angle:2007uj}
and the solid black line is 
the SIMPLE constraint on the proton cross section \cite{Felizardo:2010mi}.}
\end{figure}

\begin{figure}[t]
\includegraphics[width=17.0cm]{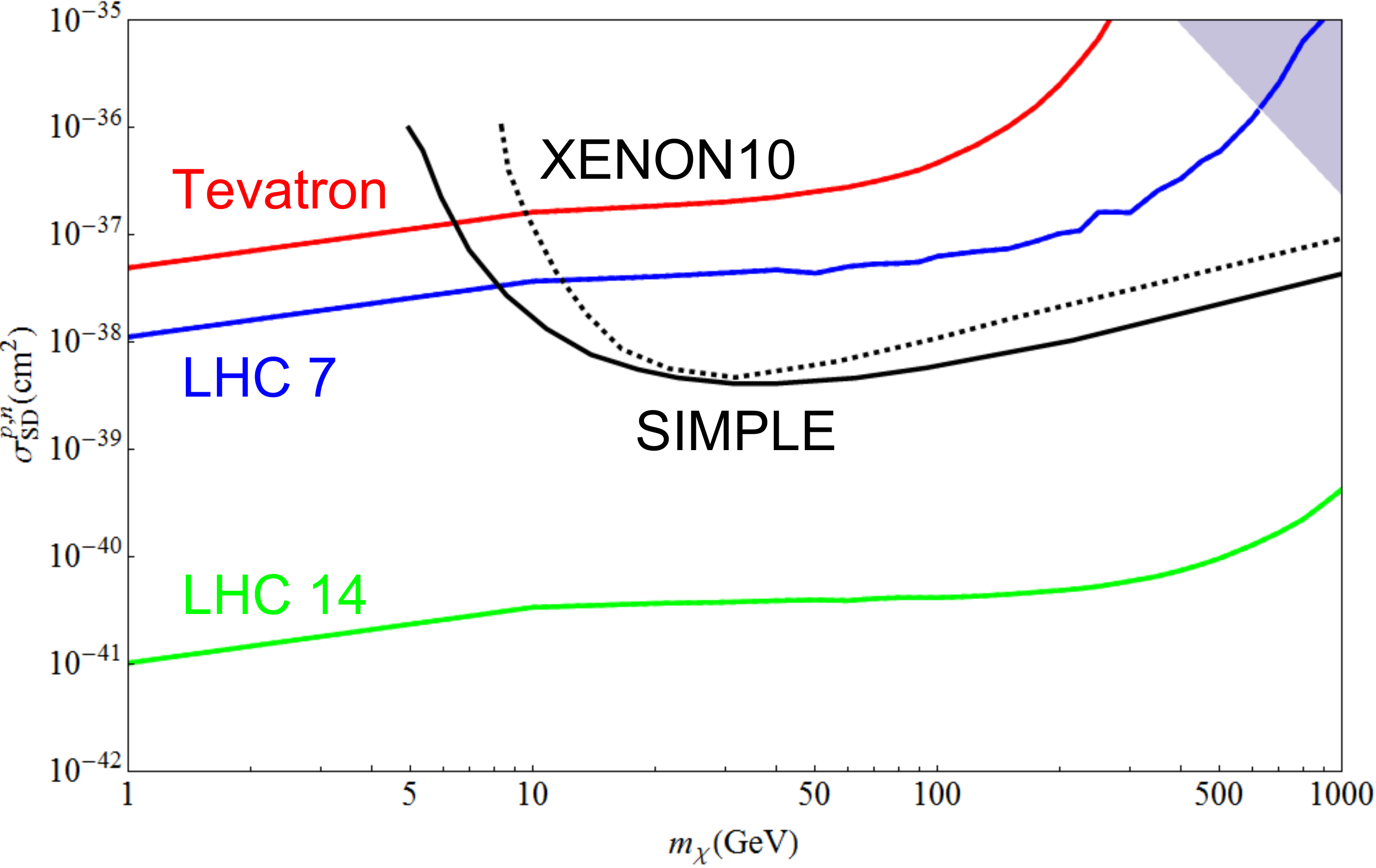}
\caption{\label{fig:dsdd} 
The same as Fig. \ref{fig:dsdu} but for the down-type quark coupling.}
\end{figure}

\begin{figure}[t]
\includegraphics[width=17.0cm]{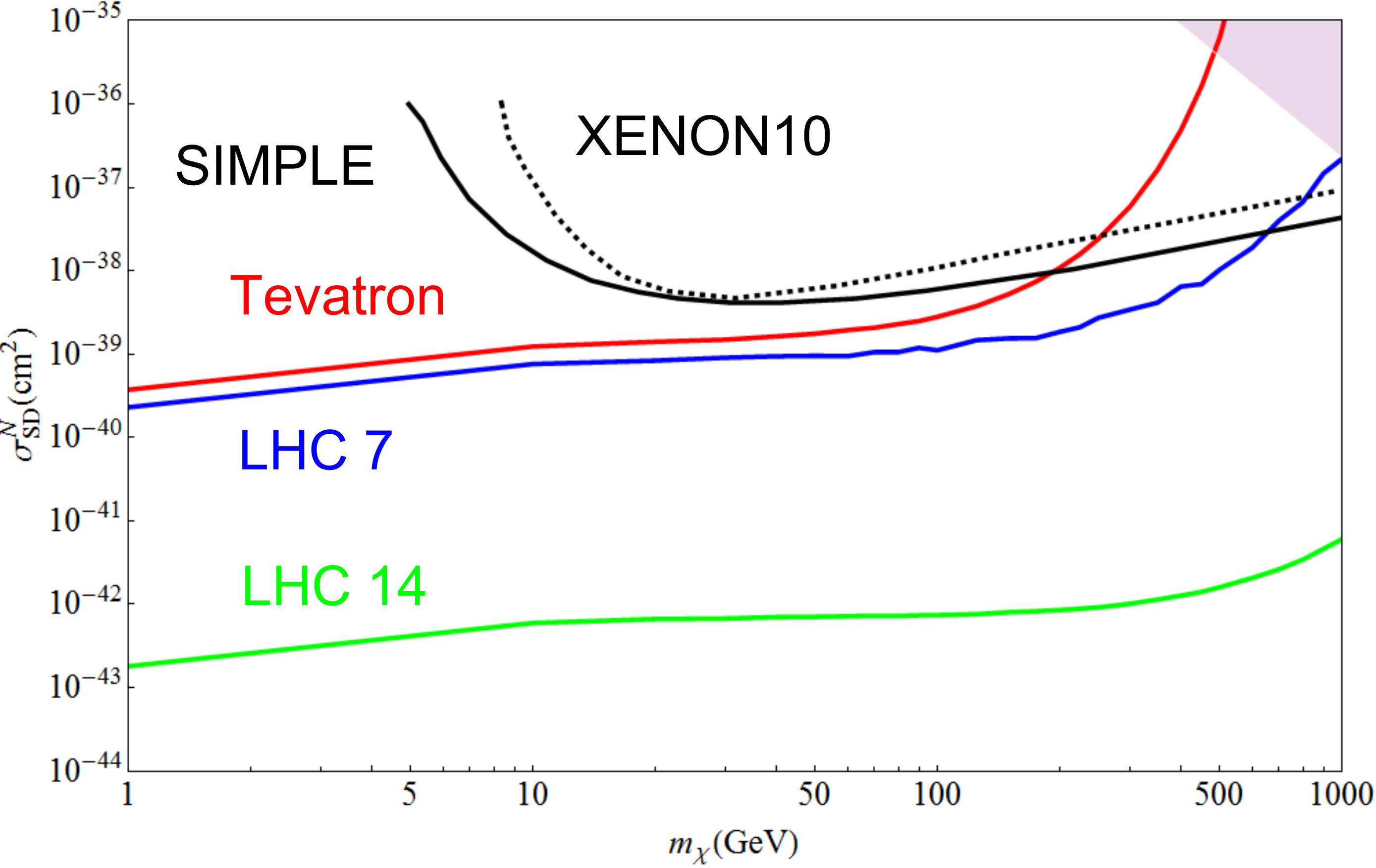}
\caption{\label{fig:dsdl1} 
Spin dependent nucleon coupling cross section assuming equal down and up type 
couplings. The red and blue lines are the constraints from the Tevatron search and 
7 TeV LHC search. The green line is the 14 TeV LHC discovery reach. The 
dashed black line is the XENON10 constraint on the neutron cross 
section\,\cite{Angle:2007uj}, the solid black line is the 
SIMPLE constraint on the proton cross section.\cite{Felizardo:2010mi}}
\end{figure}

\begin{figure}[t]
\includegraphics[width=17.0cm]{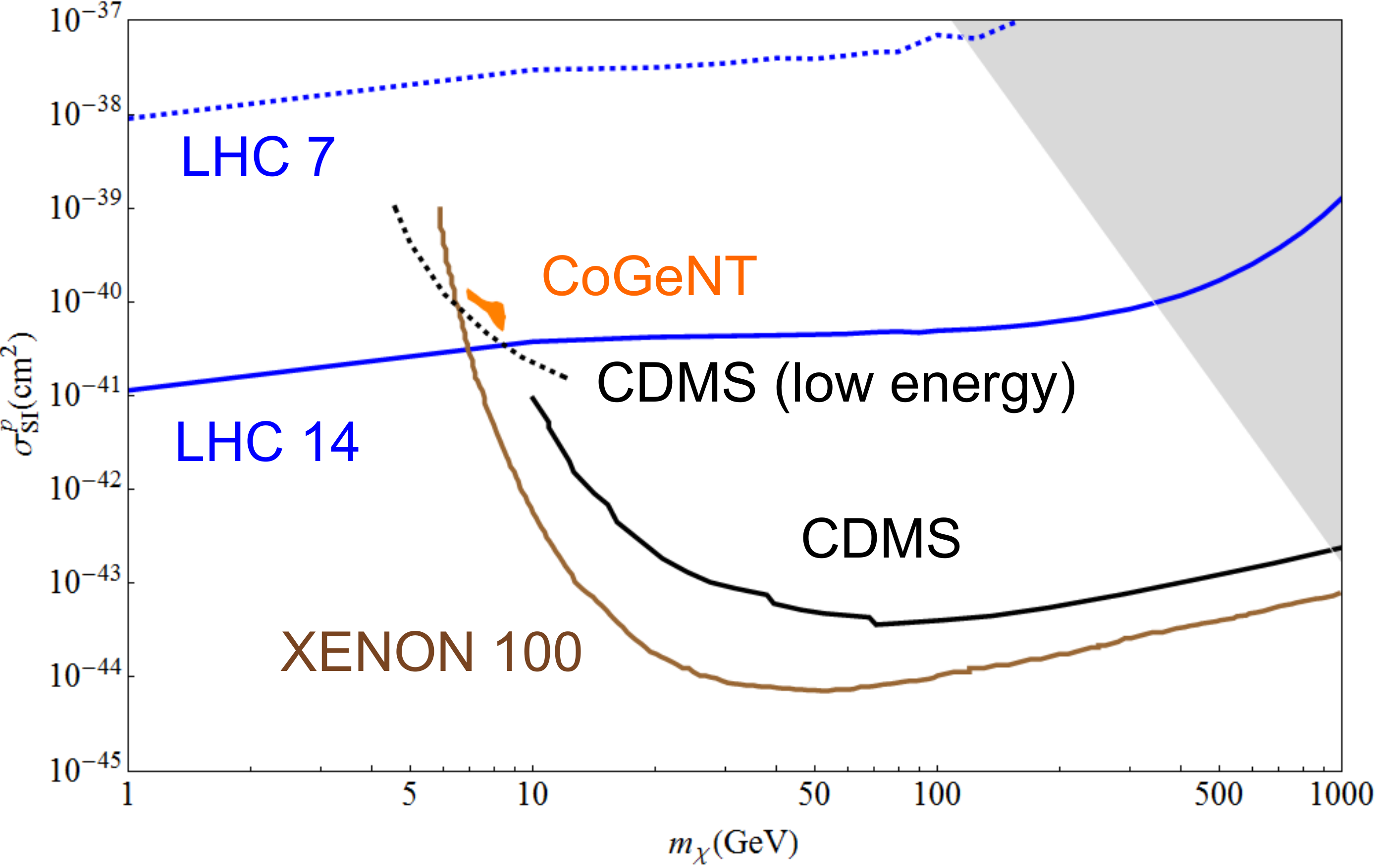}
\caption{\label{fig:dsiu} 
Spin independent proton scattering cross section assuming only up-type quark 
coupling. The red line is the constraint from the Tevatron search. The blue lines are 
the LHC 7 TeV constraint and LHC 14 discovery reach, which are dashed and solid 
respectively. The brown line is the XENON100 constraint \cite{Aprile:2010um}. The 
black lines (both solid and dashed) are the CDMS 
constraints \cite{Ahmed:2009zw,Ahmed:2010wy}. 
The orange region is CoGeNT favored results.\cite{Aalseth:2010vx}}
\end{figure}

\begin{figure}[t]
\includegraphics[width=17.0cm]{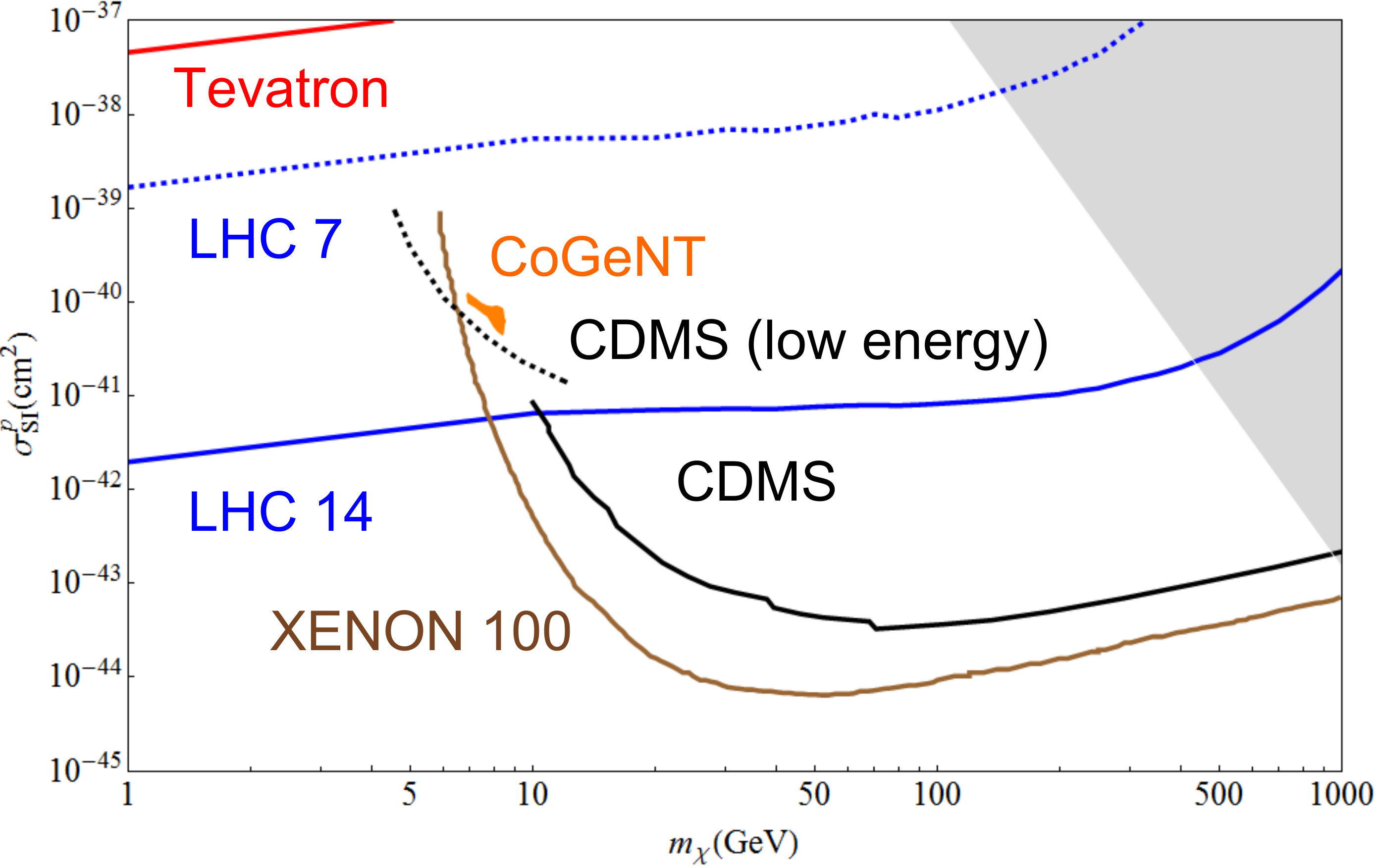}
\caption{\label{fig:dsid} 
The same as Fig. \ref{fig:dsiu} but for the down-type coupling.}
\end{figure}

\begin{figure}[t]
\includegraphics[width=17.0cm]{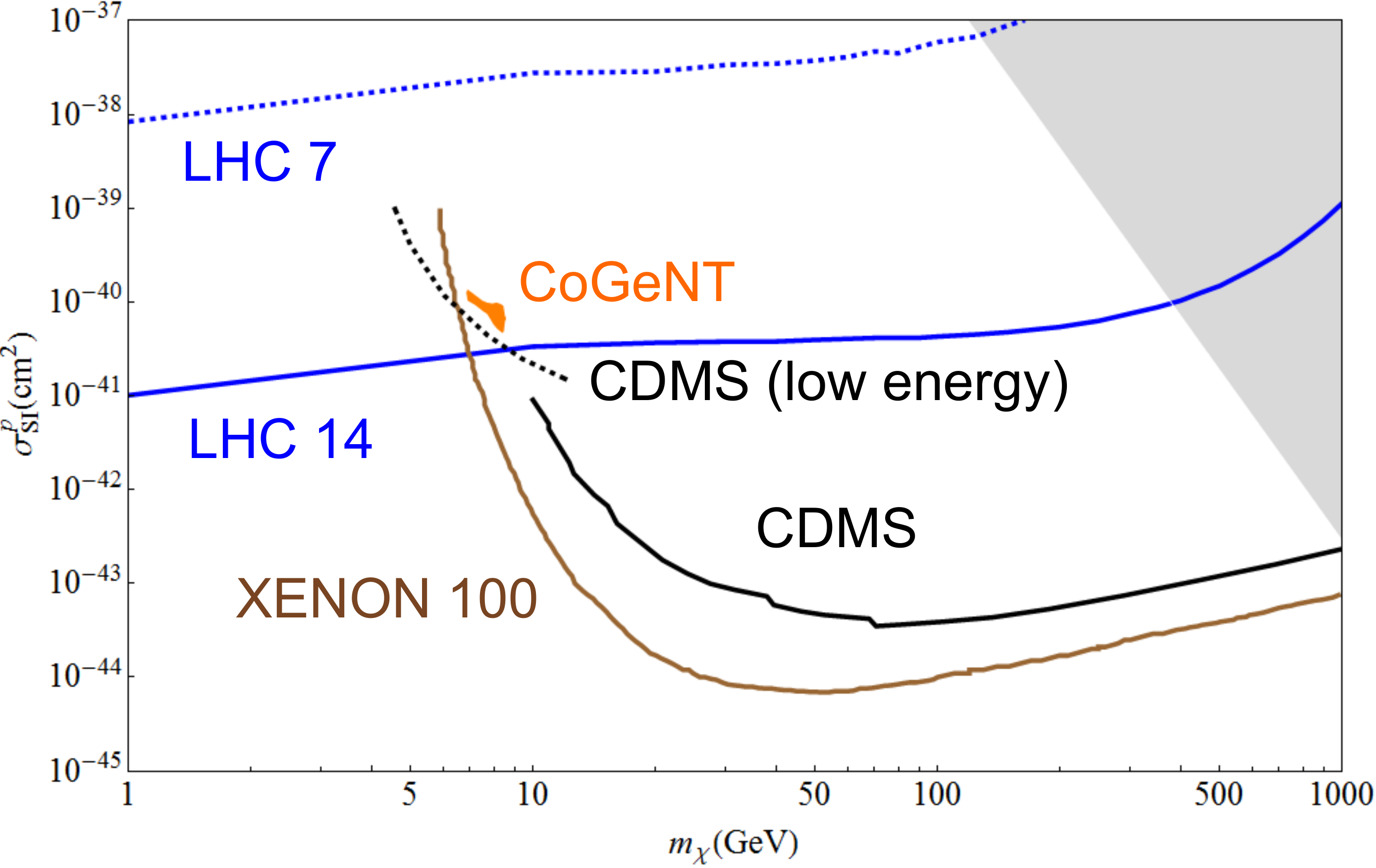}
\caption{\label{fig:dsil1} 
Spin independent coupling assuming both down and up type couplings such that the proton and neutron coupling is equal. The red line is the constraints from the Tevatron search. The blue lines are the LHC 7 TeV constraint and LHC 14 discovery reach, which are dashed and solid respectively. The brown line is the XENON100 constraint.\cite{Aprile:2010um} The black lines (both solid and dashed) are the CDMS constraints.\cite{Ahmed:2009zw,Ahmed:2010wy} The orange region is CoGeNT favored results.\cite{Aalseth:2010vx}}
\end{figure}

\begin{figure}[t]
\includegraphics[width=17.0cm]{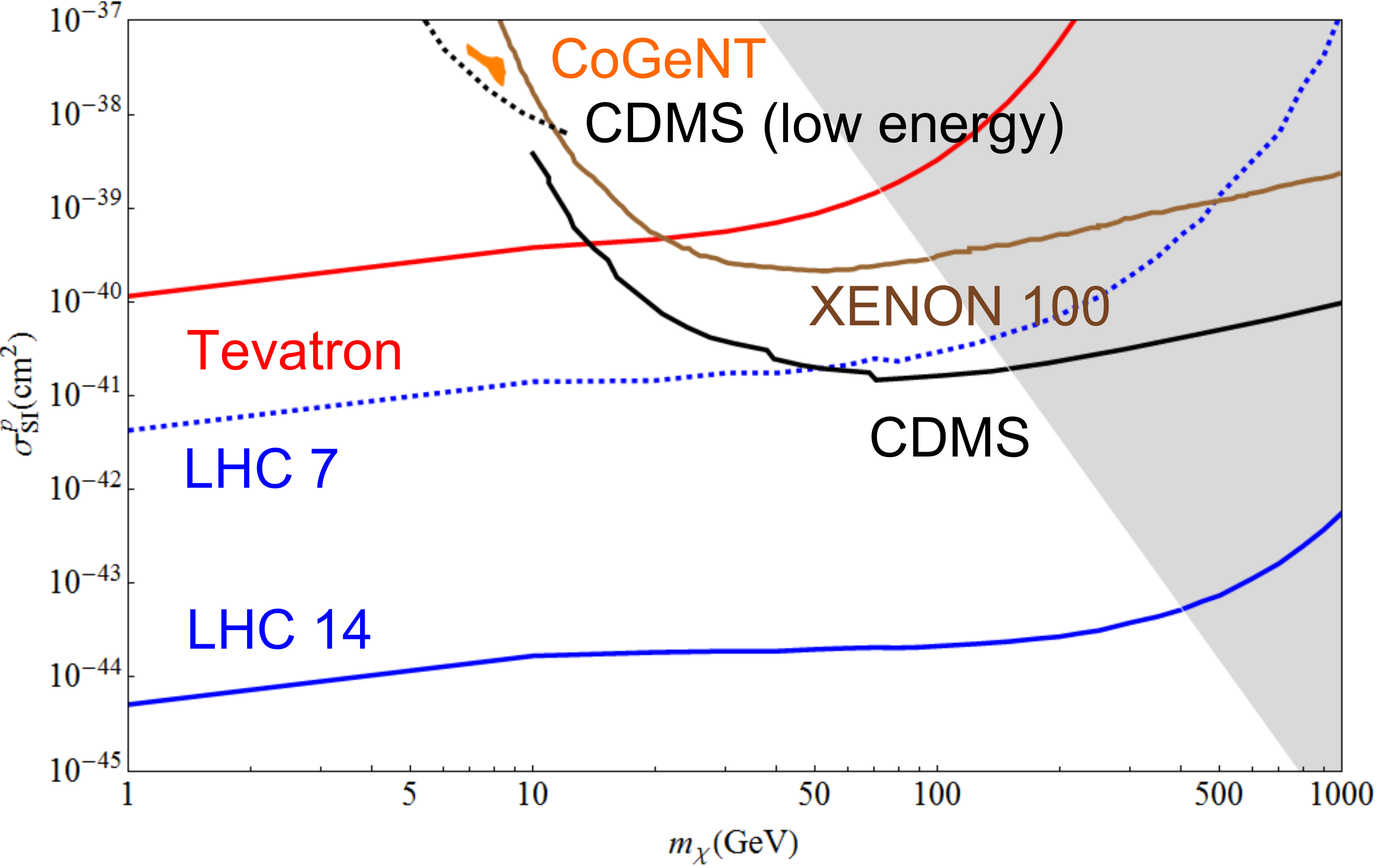}
\caption{\label{fig:dsil7} 
Spin independent coupling assuming both down and up type coupling such that the neutron to proton coupling ratio is -0.7. The red line is the constraint from the Tevatron search. The blue lines are the LHC 7 TeV constraint and LHC 14 discovery reach, which are dashed and solid respectively. The green line is the XENON100 constraint.\cite{Aprile:2010um} The black lines (both solid and dashed) are the CDMS constraints.\cite{Ahmed:2009zw,Ahmed:2010wy} The orange region is CoGeNT favored results.\cite{Aalseth:2010vx}}
\end{figure}

We translate collider bounds into limits on
spin-dependent cross sections in Figures \ref{fig:dsdu}--\ref{fig:dsdl1} 
for the cases where only the operator M6u is present, the case where
only the operator M6d is present, and the case where M6u and M6d have equal 
couplings. 
The spin independent bounds are shown on Figures \ref{fig:dsiu}-\ref{fig:dsil7}. The 
proton scattering cross section bounds for only operators M1u or M1d are plotted in 
Figure \ref{fig:dsiu} and Figure \ref{fig:dsid}, while Figure \ref{fig:dsil1} shows the 
bounds assuming both M1u and M1d are present and weighted such that the 
coupling to the proton and neutron are equal.  In Figure~\ref{fig:dsil7}, we
show bounds for $\lambda_n / \lambda_p = -0.7$, the
central value for isospin violating couplings which reconcile CoGeNT and Xenon.

\section{Conclusions}
\label{sec:conclusion}

We have extended previous studies of collider constraints on dark matter to
include isospin-violating effects and updated them to make use of the recent
null searches for jets plus missing energy based on 1 fb$^{-1}$ of LHC data.
Our effective theory description is structured based on MFV
to ensure consistency with flavor physics
observables and remain as model-independent as possible. In
particular, it faithfully reproduces the physics when
the particles mediating interactions between
dark matter and the SM are significantly heavier than the dark matter particle.
We find results which are qualitatively similar to
(though quantitatively stronger than) our previous results, with collider
limits being the strongest on models of very light dark matter and losing
sensitivity as the mass of dark matter approaches the typical energy of
collisions at the collider. 

Collider constraints on spin-dependent scattering can be appreciably weakened
by isospin violation in the UV couplings of dark matter to quarks. Suppressing
the coupling to one type of quarks does not significantly change the production
cross section at colliders for dark matter pairs, but it does remove
destructive interference in the direct detection scattering cross section,
leading to weaker limits from direct detection searches 
than for isospin conserving cases.

The effects of isospin violation in the spin-independent sector can either
strengthen or weaken collider bounds. Suppressing couplings to the heavier 
down-type quarks significantly decreases the cross section at colliders for
mass-suppressed operators, which are the main contributor to spin-independent
scattering. However, taking the preferred value for isospin violation which
allows CoGeNT to be consistent with Xenon 100 results strengthens collider
bounds considerably, as it leads to large destructive interference even within
a single nucleon as compared to the usual case of isospin conservation.
The bounds derived from colliders in this region of parameter space are not
only stronger relative to the weakened direct detection experiments, but also
stronger in the absolute sense by orders of magnitude. 7 TeV LHC results are
already competetive with the strongest direct detection bounds through a large
range of dark matter mass in this case, and future LHC reach is better up to
masses beyond 1 TeV.

These results are sensitive to the assumption that the particle mediating the
dark matter-SM interactions is heavy, and also to the assumption that such
interactions obey the MFV hypothesis. In models which predict light
mediators or more complicated flavor structures for these interactions those
effects need to be taken into account directly, either through using a UV
complete description of the dark matter scattering or altering the ratios of
couplings between the generations away from the MFV assumptions. Our results
indicate that any theory of dark matter which uses the paradigm of isospin
violation to reconcile the CoGeNT and Xenon results must either have a
collider-accessible mediator responsible for dark matter-SM interactions or
have more complicated flavor structure in its couplings. In particular,
theories which only couple the dark matter to up and down quarks, and not
members of the other generations, are much more difficult to probe at colliders
if they interact through mass-suppressed operators.

 \section*{Acknowledgements}
We thank D. Sanford for helpful conversations.  
The work of AR is supported in part by NSF grant PHY-0653656. AR would also like
to thank the Aspen Center of Physics, where part of this work was completed, for
hospitality. The work of TT is
supported
in part by NSF grant PHY-0970171 and he gratefully acknowledges the 
hospitality of the SLAC theory group, where part of this work was completed.
TT and WS acknowledge the hospitality of TASI-2011
at the University of Colorado, where some of this work was undertaken.

\end{document}